\documentclass[a4paper,11pt]{article}

\usepackage{jinstpub} 
\usepackage{lineno}
\usepackage{caption} 
\usepackage{subcaption} 

\nolinenumbers

\title{Intrinsic Energy Resolution of Plastic Scintillator Tiles: Muon Beamtest Results}

\author[a]{S. Põldmaa}
\author[b]{R. Paabo}
\author[a]{M. Rannut}
\author[b]{V. Jürgens}
\author[c]{A. Jakobson}
\author[d]{M. Schwinzerl}
\author[d]{Ş. Esen}
\author[d]{M. Van Dijk}
\author[d]{J. Petersen}
\author[d]{S. M. Zoechling}
\author[d]{M. Joos}

\affiliation[a]{Tallinn Secondary School of Science,\\ Estonia pst 6, Tallinn, Estonia}
\affiliation[b]{Hugo Treffner Gymnasium, \\Munga 12, Tartu, Estonia}
\affiliation[c]{Nõo Gymnasium, \\ K. Aigro 5, Nõo, Estonia}
\affiliation[d]{CERN,\\ Esplanade des Particules 1, Geneva, Switzerland}

\emailAdd{TBD@cern.ch}

\abstract{This study explores the feasibility of using scintillation detectors for muon calorimetry through experiments conducted at CERN's T10 beamline. Results from organic scintillators with varying thicknesses and readout methods, as well as a lead-glass calorimeter, showed minimal correlation between muon energy and the peak or spread of the detected signal amplitudes. These findings highlight the influence of temperature on SiPMs and indicate a significant amount of contamination in the muon beam.}


\begin{document}
\maketitle
\flushbottom

\section{Introduction}
The energy spectrum of cosmic ray muons has been well studied at ground level, but data collected from higher altitudes remains insufficient. This is largely due to the inherent limitations of high-altitude balloon flights, where the weight of the equipment must be kept minimal~\cite{balloons}.

For this purpose, plastic scintillators coupled to SiPMs are well-suited, thanks to their lightweight nature. While such setups can be employed in high-energy muon calorimetry, where bremsstrahlung dominates~\cite{bremmstrahlung}, their application in cosmic ray muon calorimetry is hindered by the statistical nature of ionization~\cite{daniela_esitlus} and the minimal variations in stopping power that occur in the energy range of cosmic ray muons~\cite{StoppingPower, PDG_Bethe}.

Additionally, the probabilistic nature of ionization is especially pronounced when using thinner scintillators, as the absorber length far subceeds the muon range~\cite{daniela_esitlus}. Furthermore, the non-proportionality is exacerbated by the nonlinear response between emitted light and the SiPM signal~\cite{moszynski_sipm}.

\begin{figure}[htbp] \centering \includegraphics[width=0.8\textwidth]{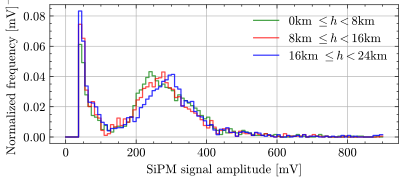} \caption{Normalized histograms of the COSMOSS detector's voltage amplitudes recorded in various altitude ranges during the high-altitude balloon flight. \label{fig:results}} \end{figure}

Despite these shortcomings, experiments conducted with the COSMOSS detector (COSt-efficient Muon Observation with Scintillators and SiPMs), onboard a high-altitude balloon flight revealed contrary findings. During the flight, the SiPM signal amplitude was observed to increase with altitude, as illustrated in Figure~\ref{fig:results}. This increase was attributed to the higher energy of muons at greater altitudes, which raises ionization rates according to the Bethe-Bloch equation. These results suggest a correlation between muon energy and SiPM signal amplitude, opening new avenues for muon calorimetry.

To test the energy resolution of the COSMOSS detector, along with other scintillator-based detectors and a lead-glass calorimeter, a beamtest was carried out in September 2024 at the T10 beamline in the East Area~\cite{cern_ps} of the Proton Synchrotron at CERN.

\section{Stratospheric flight results}
\begin{figure}[htbp]
\centering
\begin{subfigure}[b]{0.7\textwidth}
    \includegraphics[width=\textwidth]{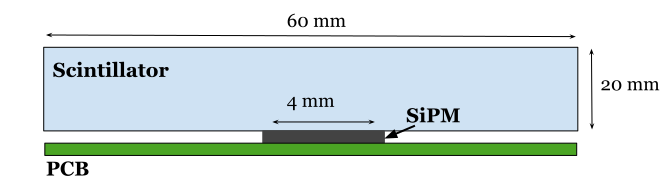}
    \caption{}
\end{subfigure}
\hfill
\begin{subfigure}[b]{0.7\textwidth}
    \includegraphics[width=\textwidth]{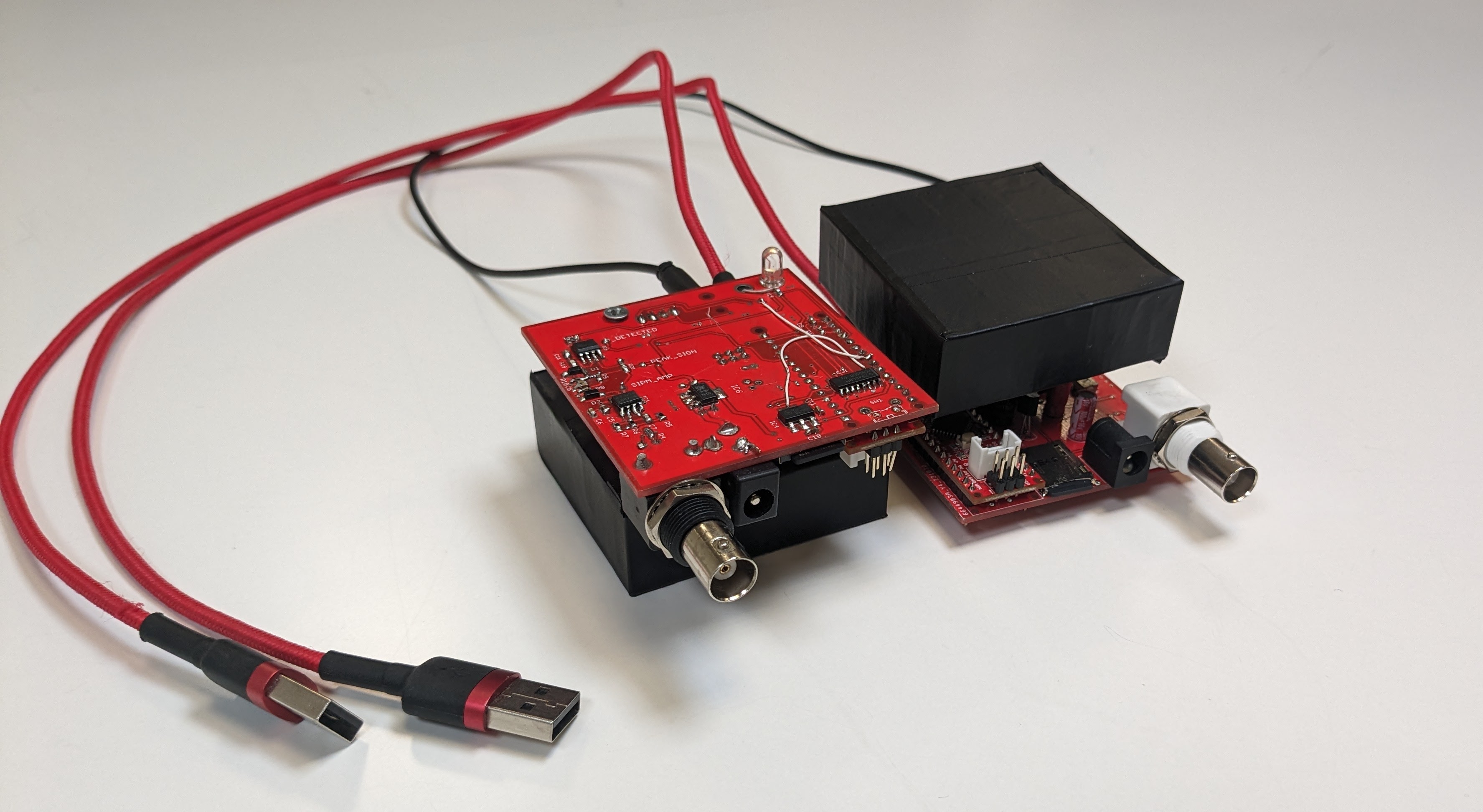}
    \caption{}
\end{subfigure}
\caption{a) Layout of the COSMOSS detector b) Photo of two COSMOSS detectors, which were set into coincidence during the high-altitude balloon flight. \label{fig:produkt}}
\end{figure}

The COSMOSS detector (shown on Figure~\ref{fig:produkt}) consisted of a polystyrene scintillator (Advantech's UPS-89~\cite{scintillator}) with a central emission wavelength of $\lambda = 418 $ nm and an SiPM. To the geometric centre of the $6 \times 6$ cm scintillator tile of thickness $2$ cm and, an SiPM (Hamamatsu S14161-4050HS-06~\cite{Hamamatsu}) was coupled using optical gel EJ-550~\cite{gel}, as illustrated on Figure~\ref{fig:produkt}. The scintillator tile was wrapped in aluminum foil for better reflectivity, and then covered with electrical tape, ensuring light-tightness. 

The electronics of the COSMOSS detector were inspired by the popular CosmicWatch detector~\cite{CWD}, but were modified to accommodate the higher bias voltage of the Hamamatsu S14161-4050HS-06~\cite{Hamamatsu} SiPM, and added was a comparator. The SiPM signal was passed through an amplifier and a peak detection circuit, and was then read out by an Arduino Nano. For each coincidence event, recorded were the event's timestamp and pulse height as an ADC value. From the ADC value, a voltage on the SiPM was extrapolated from calibration results. 

\begin{figure}[htbp]
\centering
\includegraphics[width=0.5\textwidth]{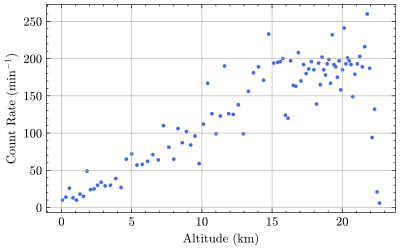}
\caption{Measured count rates at various altitudes.  \label{fig:cpmvsh}}
\end{figure}

Onboard the high-altitude balloon flight, two  detectors were set in coincidence, while a vertical distance of $5$ cm was kept between the two scintillator tiles, in order to reduce the observable solid angle. The ascent of the balloon lasted for 86 minutes, and the balloon reached a maximum height of 24 km. The count rates of coincidence events at various altitudes are shown on Figure~\ref{fig:cpmvsh}, and the recorded voltage histograms on Figure~\ref{fig:results}.

\section{Beam Line and Detectors}
\begin{figure}[htbp]
\centering
\includegraphics[width=\textwidth]{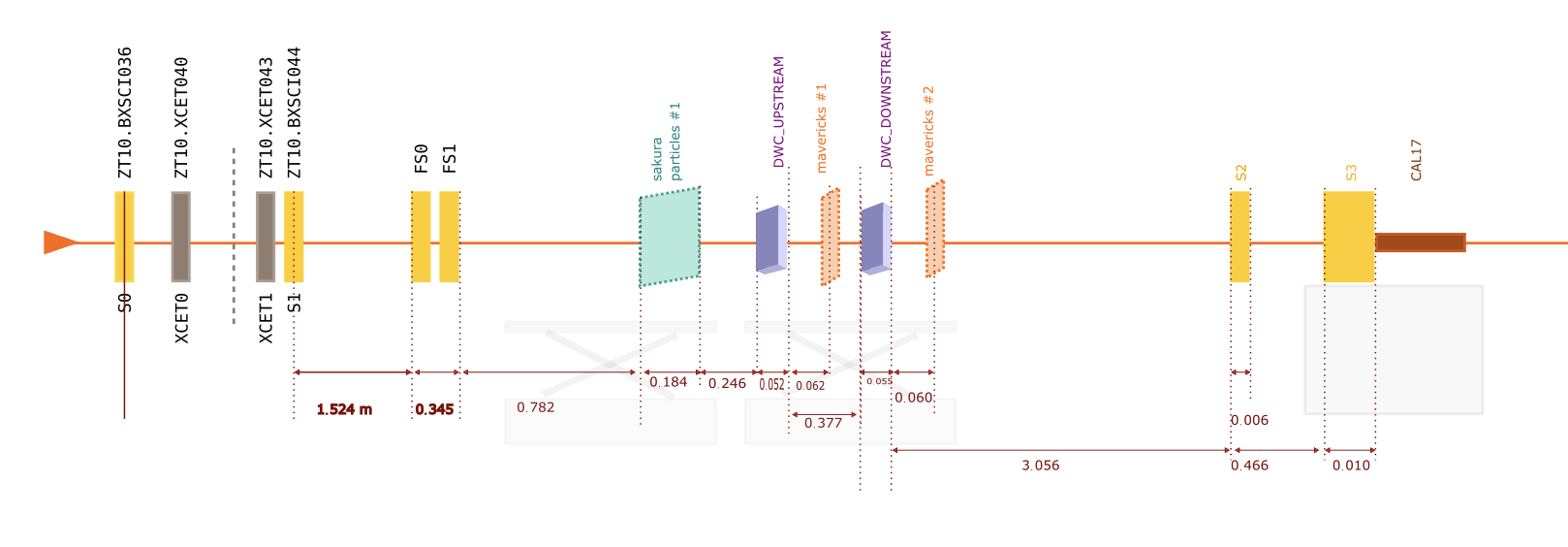}
\caption{Layout of the experiment.  Mavericks \#1 refers to the COSMOSS detector. \label{fig:composition}}
\end{figure}

The components of the experimental setup are shown schematically in Figure~\ref{fig:composition}. The DAQ triggered on the signal S0\&S1\&S2\&S3\&NOT BUSY, where NOT BUSY was a busy signal from the oscilloscope. The trigger signal featured both the upstream (S0, S1) and downstream (S2, S3) scintillators, to allow for easier timing-in of signals.

The used scintillation detectors included the COSMOSS detector and two additional polystyrene scintillators, designated as S2 and S3. S2 and S3 differed in width: S2 had a width $0.6$ cm, whereas S3 was $10$ cm wide. In addition to the scintillation detectors, the energy resolution was also studied in a lead-glass calorimeter (CAL) of width $30$ cm.  

The detectors were aligned to an accuracy of $1$ mm using a laser cross. A negative beam was used, from which a muon beam was generated using a beam stopper. It should be noted that the beam stopper was unable to stop all hadrons in the primary beam, resulting in a significant amount of contamination in the muon beam, primarily in the form of pions.

\begin{figure}[htbp]
\centering
\includegraphics[width=\textwidth]{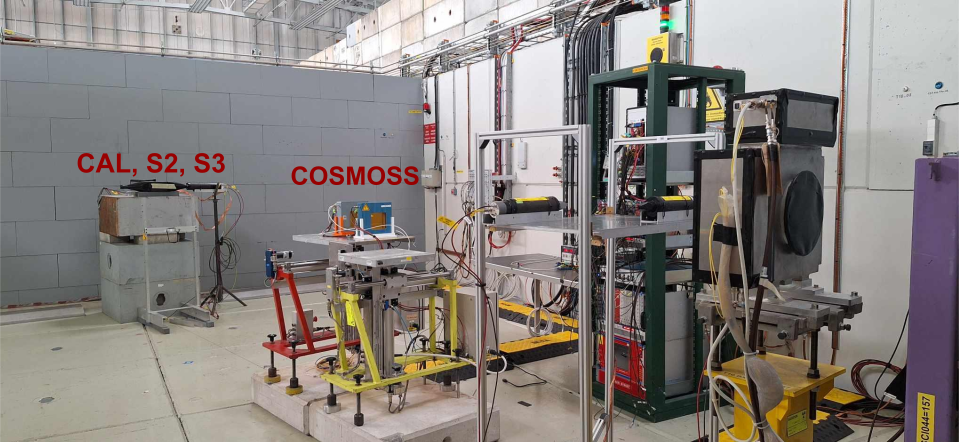}
\caption{Testbeam setup. \label{fig:labelled}}
\end{figure}

The bias voltage on the COSMOSS detector's SiPM was consistently set to $40.25$ V, which is $2.25$ V above the breakdown voltage. The SiPM was read out by directly probing the SiPM's signal line by an 8-channel Tektronix MSO58LP oscilloscope ~\cite{oscilloscope}. The oscilloscope was set to a sampling rate of $1.5625$ GS/s, and the sample length was set to $8$k samples, corresponding to a sample window of $5\ \mu s$. This window was significantly longer than required to capture the full duration of the SiPM signal; however, it was chosen to accommodate other detectors connected to the oscilloscope, which are not included in this study.

Both the scintillators S2 and S3, and CAL were coupled to PMTs using light guides, and their readout was performed with QDCs. The used PMTs (Thorton EMI 9813 KA) featured a maximum quantum efficiency of 25\% for a wavelength of $k = 380$nm. The analog signals from the S0 and S1 PMTs were fed into LeCroy 620AL NIM discriminator units ~\cite{discriminator} with a threshold of $-30$ mV and signal length of $100$ ns. The oscilloscope's busy signal was emitted by a Raspberry Pi as a +3.3V TTL signal, which was then fed into the same discriminator unit with a disabled monostable behaviour, and a threshold of $2$ V.

The two XCETs, designated XCET040 and XCET043 in Figure~\ref{fig:composition}, were filled with $CO_2$ gas. The pressure adjustment was done remotely through a gas rack, and the pressure was brought maximally up to $4.1$ bar in the XCET043, and to $14.4$ bar in the XCET040. The light was collected using PMTs, which were read out using discriminator units.

\section{Momentum bracketing}
In order to differentiate muons of specific momenta, a novel technique of momentum tagging was implemented, using two threshold Cherenkov counters (XCETs~\cite{xcet}). One of the XCETs was constantly set to a slightly higher pressure –– this allowed to set two, slightly differing threshold momenta. A particle which emitted light in only one of the detectors was known to lie in the selected momentum range, i.e. in the momentum bracket.

The refractive index $n$ of the gas in the XCET scales linearly with pressure $p_{TH}$ as the following:
\begin{equation}
\label{eq:refrindex}
n = 1 + kP[\mathrm{Bar}] \,,
\end{equation}
where $k$ is a gas-specific constant. As a result of the emission spectrum of XCETs and their efficiency as a function of wavelength~\cite{mirror}, XCETs become most sensitive in the UV range ~\cite{xcet_efficiencies}, where for $CO_2$, k takes the value $k = 4.50 \cdot 10^{-4} \, \mathrm{bar^{-1}}$~\cite{CO2} at $250$ nm.

The threshold energy $E$ in an XCET at pressure $p_{TH}$ was therefore
\begin{equation}
\label{eq:energy}
E = \left( \frac{1+kp_{TH}}{\sqrt{(kp_{TH})^2+2kp_{TH}}}-1 \right)mc^2 \,.
\end{equation}

According to the Frank-Tamm relation, a particle with momentum only slightly above the threshold might go undetected in an XCET, as too few photons may be emitted for detection~\cite{neutrino}. This effect is further exacerbated by chromatic dispersion, which causes the true threshold energies to deviate from those calculated in ~\ref{eq:energy}. To account for these effects, pressure corrections were needed. 

The pressure corrections were determined through pressure scans and verified using Monte Carlo simulations in GEANT4. It was found that physical pressures $p_{40}$ and $p_{43}$ in XCET040 and XCET043 respectively, correspond to the theoretical pressures

\begin{equation}
\label{eq:correction}
\begin{aligned}
p_{TH} &= 0.975048 \times p_{40}[\mathrm{Bar}] - 0.153522 \ \mathrm{Bar} \,,
\\
p_{TH} &= 0.922359 \times p_{43}[\mathrm{Bar}] - 0.022896\ \mathrm{Bar} \,.
\end{aligned}
\end{equation}
where $p_{TH}$ is the pressure which was used for the calculation~\ref{eq:energy}.

The results of the GEANT4 simulation agreed roughly with the aforementioned pressure corrections, suggesting a correction of $0.30$ Bar for both XCETs, regardless of the pressure. The pressure corrections specified in~\ref{eq:correction} were applied to set the pressure in the XCETs to achieve the desired threshold energies. At higher beam energies, the energy bracket was widened to compensate for the reduced muon count rates. In the following discussion, the bracket's energy is defined as the midpoint of the bracket, with error bars representing its width.

\section{Signal analysis}
The SiPM pulse as recorded by the oscilloscope, was analyzed in two ways: by the amplitude of the signal, and by the area under it. 

\begin{figure}[htbp]
\centering
\includegraphics[width=.5\textwidth]{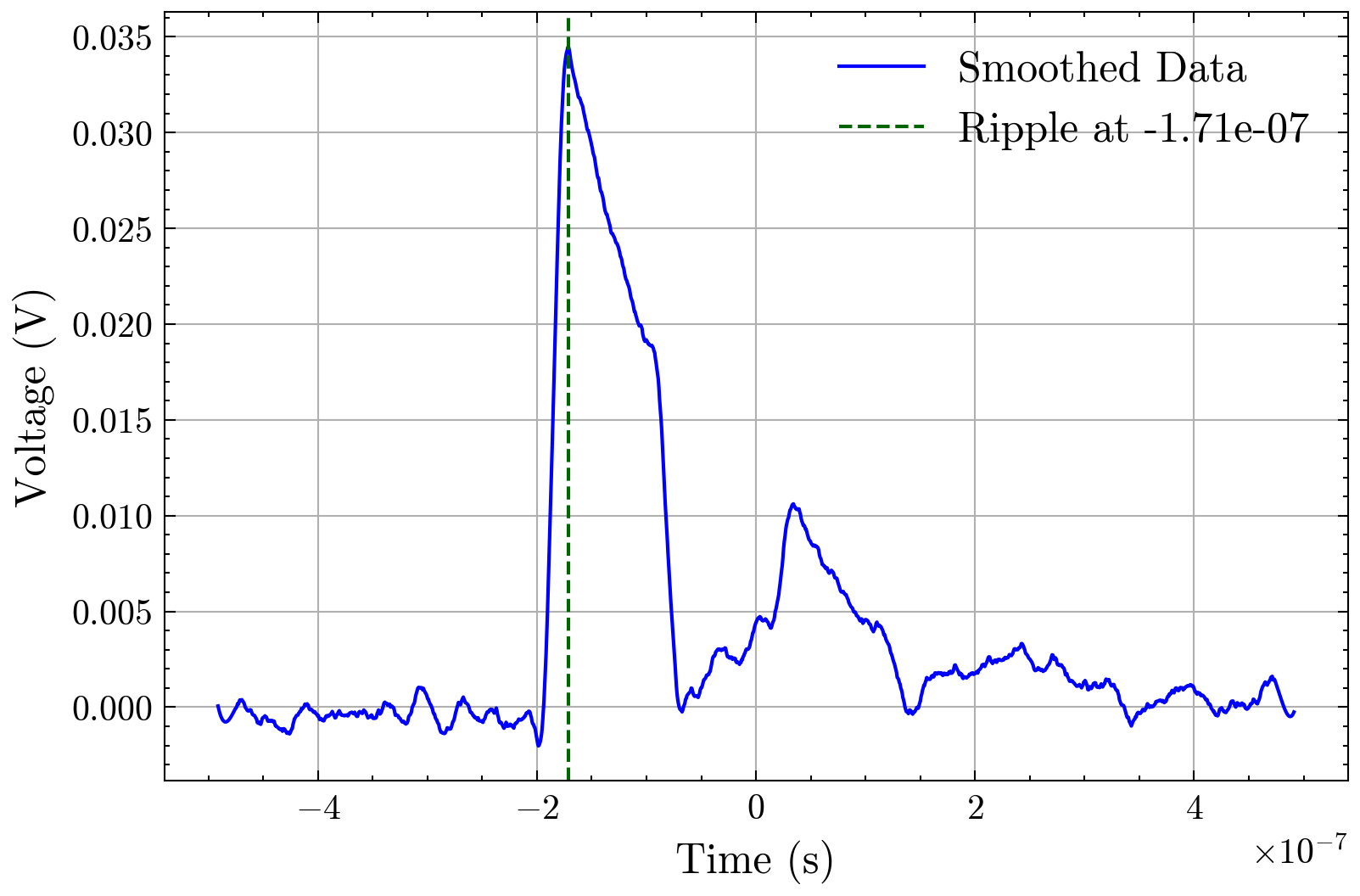}
\caption{Afterpulse observed on the SiPM. \label{fig:afterpulse}}
\end{figure}

The readout system introduced a power supply ripple of 1 MHz. The effect of this ripple was mitigated both through data analysis, where entries with this characteristic frequency were discarded, and by powering the detector with an external power supply instead of directly using the electrical grid. In addition, the SiPM exhibited very strong afterpulsing, an example of which is shown on Figure~\ref{fig:afterpulse}.

According to ~\cite{afterpulsing}, the two factors affecting the probability of afterpulsing are bias voltage and SiPM cell size. The bias voltage was decreased to $40.25$ V in an attempt to mitigate the effect of afterpulsing, however this did not resolve the issue. When analyzing the signal amplitude, afterpulses were ignored. However, when calculating the area under the signal, both the primary pulse and afterpulse were recorded as a single event.

Both the SiPM signal amplitudes and the QDC data from S2 and S3 followed Landau-convoluted Gaussian distributions, as was also observed by previous users of the beamline ~\cite{convolution}. The  distributions were analyzed through the MPV and Landau sigma parameters, which were found from ROOT's Langaus fit \cite{langaus}.

\section{Results}

\begin{figure}[htbp]
\centering
\begin{subfigure}[b]{0.45\textwidth}
    \includegraphics[width=\textwidth]{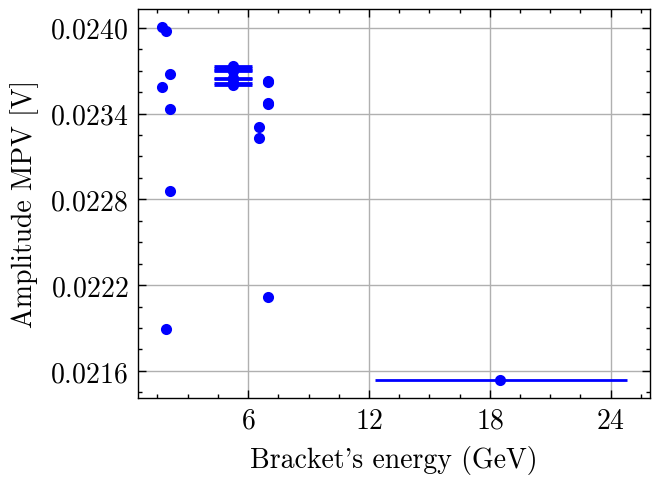}
    \caption{MPV}
\end{subfigure}
\hfill
\begin{subfigure}[b]{0.45\textwidth}
    \includegraphics[width=\textwidth]{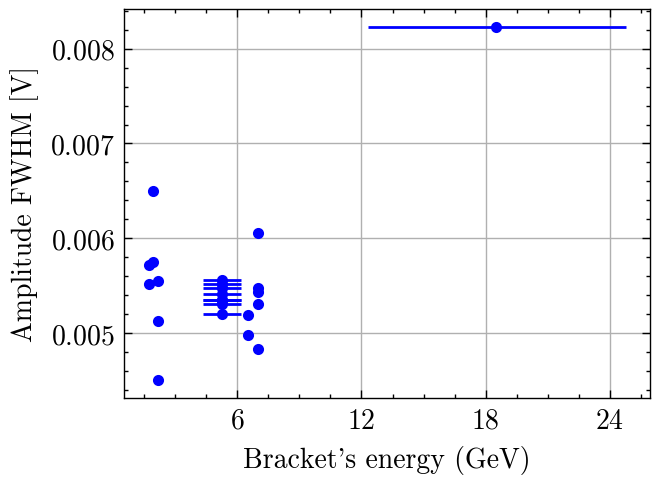}
    \caption{Landau sigma}
\end{subfigure}
\caption{The MPVs and Landau sigmas of observed signal amplitude distributions at various muon energies. \label{fig:amplitudes}}
\end{figure}

The effects of muon energy on the COSMOSS detector's signal amplitude distributions are shown in Figure~\ref{fig:amplitudes}, where the error bars represent the width of the energy bracket. The Pearson correlation coefficients between muon energy and the amplitude histogram's MPV were found to be -0.027, and -0.001 for the Landau sigma. These results indicate a lack of significant correlation between muon energy and signal amplitude, suggesting that muon energy does not significantly influence the SiPM's signal output and, consequently, has little effect on the energy deposited in the scintillator.

\begin{figure}[htbp]
\centering
\begin{subfigure}[b]{0.45\textwidth}
    \includegraphics[width=\textwidth]{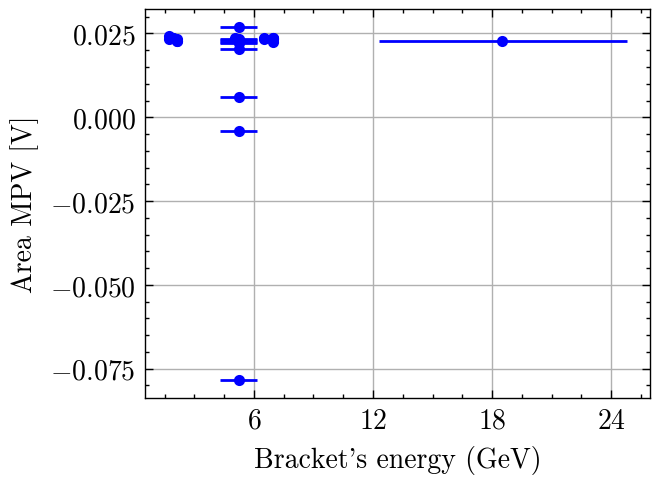}
    \caption{MPV}
\end{subfigure}
\hfill
\begin{subfigure}[b]{0.45\textwidth}
    \includegraphics[width=\textwidth]{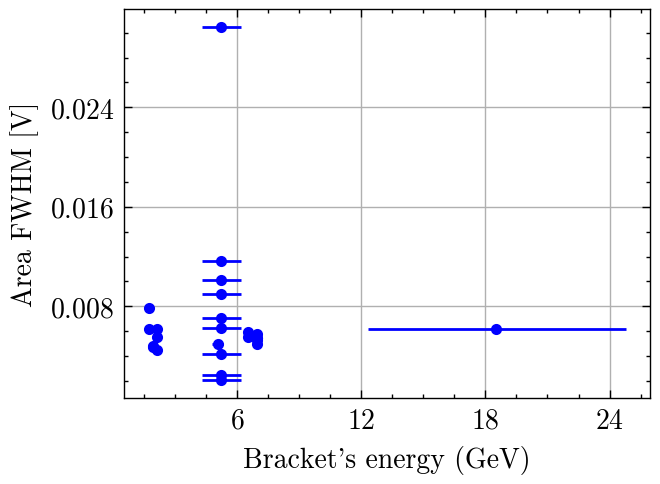}
    \caption{Landau sigma}
\end{subfigure}
\caption{The a) MPVs and b) Landau sigmas of COSMOSS detector's signal area distributions at various muon energies. \label{fig:areas}}
\end{figure}

Figure~\ref{fig:areas} illustrates the relationship between muon energy and the area of signal produced by the COSMOSS detector. In some cases, inaccuracies in determining the ground level from the oscilloscope data resulted in the area dropping below zero. The Pearson correlation coefficients for the MPV and Landau sigma were -0.001 and 0.004. Compared to signal amplitude, the signal area exhibits an even lower degree of correlation with muon energy, further confirming the lack of influence between  muon energy and SiPM response.

\begin{table}[htbp]
\centering
\caption{Correlation coefficients for MPV and Landau sigma of the signal distributions observed by various detectors.\label{tab:table}}
\smallskip
\begin{tabular}{l|c c c c c}
\hline
Parameter & COSMOSS amplitude & COSMOSS area & S2 & S3 & CAL\\
\hline
MPV & 0.129 & -0.027 & -0.206 & -0.244 & -0.044\\
Landau sigma & -0.001 & 0.004 & 0.163 & -0.057 & 0.312\\
\hline
\end{tabular}
\end{table}

Table~\ref{tab:table} presents the correlation coefficients for S2, S3, and CAL. For all these detectors, the most probable value (MPV) of the QDC data exhibited a very weak negative correlation with muon energy. The strongest correlation, 0.312, was observed for the calorimeter’s sigma value, suggesting that the QDC data distribution broadens as muon energy increases. However, this effect was not observed for any of the scintillation detectors.

\begin{figure}[htbp]
\centering
\includegraphics[width=.6\textwidth]{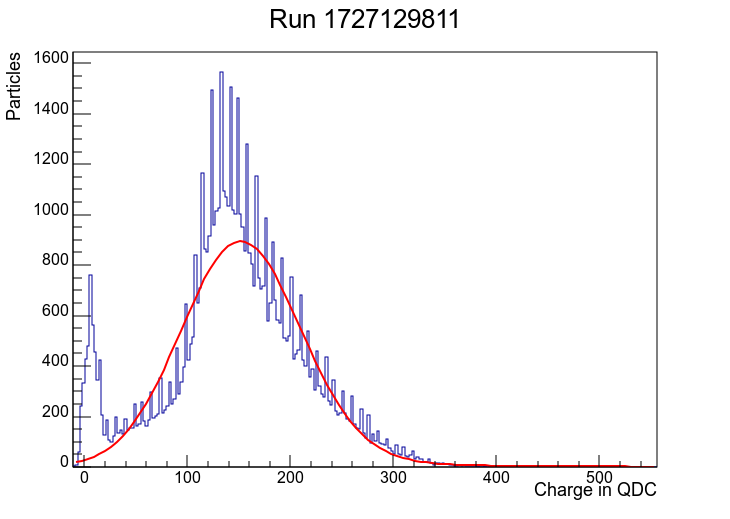}
\caption{Double-peaked histogram of QDC counts measured by CAL in an energy momentum bracket of $4.31$ -- $6.17$ GeV at 7 GeV beam energy.  \label{fig:punchthrough}}
\end{figure}

The QDC data from the calorimeter revealed significant punch-through from the beam stopper, leading to the emergence of a secondary peak at lower QDC values, as illustrated in Figure~\ref{fig:punchthrough}. The widening of the QDC spectrum at higher energy brackets suggests that the punch-through effect becomes increasingly pronounced at higher beam energies, which were used for runs with higher muon energy brackets.

\begin{figure}[htbp]
\centering
\begin{subfigure}[b]{0.45\textwidth}
    \includegraphics[width=\textwidth]{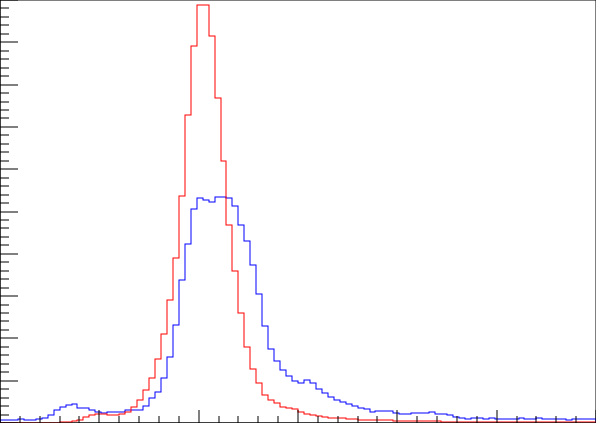}
    \caption{Signal amplitude. x-axis: Voltage ($10 mV$), y-axis: normalized frequency.}
\end{subfigure}
\hfill
\begin{subfigure}[b]{0.45\textwidth}
    \includegraphics[width=\textwidth]{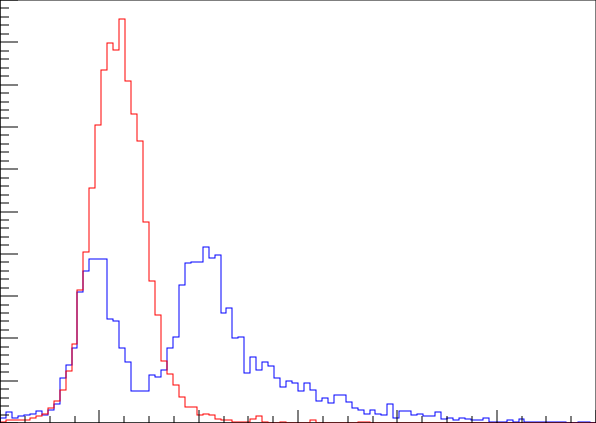}
    \caption{Signal area. x-axis: Area ($10 mV\cdot ns$), y-axis: normalized frequency.}
\end{subfigure}
\caption{Normalized histograms of COSMOSS detector's signal amplitude and area at room temperature (red) and at $-78^\circ$ (blue), at 2GeV beam energy with no momentum bracketing.\label{fig:temp}}
\end{figure}

To further investigate the potential causes of the increased signal amplitudes observed during the balloon flight, the SiPM was cooled in a dry ice-filled box to simulate the colder temperatures experienced during the  flight. The SiPM's temperature was monitored using a thermocouple throughout the cooling process, with the COSMOSS detector reaching a minimum temperature of $-78^\circ$C. Figure~\ref{fig:temp} compares the distributions of signal amplitudes and areas at room temperature and $-78^\circ$C, using the same beamfile for both runs without momentum bracketing. 

As seen in Figure~\ref{fig:temp}, the signal amplitude shifts slightly to higher voltages, with the MPV increasing from 18mV to 25mV as the SiPM cooled down, proving that the trend observed during the high-altitude balloon flight (Figure~\ref{fig:results}), had been caused by the change in temperature. 

Notably, the histogram of areas developed a second peak at $-78^{\circ}C$, suggesting that signal duration is strongly temperature-dependent. This could indicate a higher probability of afterpulsing at lower temperatures. Although previous studies~\cite{afterpulsing_constant}, have confirmed that in the temperature range of $-20^{\circ} C--60^{\circ} C$, afterpulsing probability remains constant, it has also been shown that a decfrease in temperature will lower the breakdown voltage probability~\cite{sipms}. As a result, the SiPM operates at a higher overvoltage, leading to an increased probability of afterpusling~\cite{afterpulsing}.

\section{Discussion and Conclusions}

This study investigated the feasibility of using scintillation detectors for muon calorimetry. The findings indicate no significant correlation between muon energy and the scintillator response, regardless of whether the detector was coupled to a silicon photomultiplier (SiPM) or a photomultiplier tube (PMT). However, the spread of the calorimeter's QDC data appeared to be positively correlated with muon energy, likely due to the pronounced effect of punch-through at higher beam energies.

The absence of a strong correlation between muon energy and detector response can be attributed to the probabilistic nature of energy loss, as described by the Bethe-Bloch formula. At cosmic ray muon energies, stopping power remains relatively constant, with only a slight increase at higher energies. This minor increase is overshadowed by stochastic fluctuations in stopping power among muons of the same energy, leading to significant variability in scintillation light output. Additionally, photodetection introduces further nonlinearities, particularly in SiPMs, as previously reported in the literature~\cite{moszynski_sipm}.

The results obtained from the high-altitude balloon flight with the COSMOSS detector were instead attributed to the temperature sensitivity of SiPMs. A decrease in temperature not only increased the signal amplitude but also seemed to raise the probability of afterpulsing.

In summary, while scintillation detectors are effective for muon detection, their role in calorimetry is limited due to the weak correlation between muon energy and detector response. Additionally, the temperature sensitivity of SiPM-based detectors remains a key challenge, affecting their stability and accuracy. Further research is required to better understand SiPM performance at low temperatures, particularly its impact on afterpulsing probability.

Moreover, the proposed momentum bracketing method proved to be reliable and holds great potential for future applications. However, additional studies are first necessary to evaluate the purity of muon beams in the East Area and to assess whether the inefficiency of threshold Cherenkov detectors compromises the effectiveness of this method.

\acknowledgments

The authors would like to thank the Beamline For Schools (BL4S) team as well as technical supporters of BL4S for their insights, encouragement, and immense technical support. The authors also wish to thank the CERN \& Society Foundation, without whom this work would never have been possible. And finally, we thank CERN for providing us with particle beams of excellent quality.


\bibliographystyle{JHEP}
\bibliography{biblio.bib}

\end{document}